\documentstyle[12pt]{article}
\begin{document}

\title{The generation of entangled photon states by using linear optical elements}
\author{XuBo Zou, K. Pahlke and W. Mathis \\
Electromagnetic Theory Group at THT,\\
 Department of Electrical
Engineering, University of Hannover, Germany} \maketitle
\begin{abstract}
We present a scheme to generate the polarization-entangled
two-photon state
$\frac{1}{\sqrt{2}}(|H\rangle|V\rangle+|V\rangle|H\rangle)$, which
is of much interest in the field of quantum information
processing. Furthermore we demonstrate the capability of this
concept in respect of a generalization to entangle $N$-photon
states for interferometry and lithography. This scheme requires
single-photon sources, linear optical elements and a multi-fold
coincidence detection.

 PACS number(s):03.65.Ud,42.50.-p
\end{abstract}

The generation of entangled quantum states plays a prominent role
in quantum optics. An experimental realization in this context can
be achieved with trapped ions \cite{qkc}, Cavity QED \cite{ar} or
Bose-Einstein condenses \cite{am}. Experiments with
polarization-entangled photons opened a whole field of research.
Polarization-entanglement was used to test the Bell-inequality
\cite{daa} and to implement quantum information protocols like
quantum teleportation \cite{db}, quantum dense coding \cite{km}
and quantum cryptography \cite{ds}. The experimental generation of
GHZ-states of three or four photons were reported \cite{dj}. These
polarization-entangled photon states were only been produced
randomly, since there is no way of demonstrating that polarization
entanglement was generated without measuring and destroying the
outgoing state \cite{kb}. Some quantum protocols like error
correction were designed for maximally entangled quantum states
without random entanglement \cite{dbo}. Thus, a photon source is
needed, which produces maximally polarization-entangled outgoing
photons. Remarkably, efficient quantum computation with linear
optics was put forward \cite{erg}. Such schemes can be used
directly to generate polarization-entangled quantum states. It was
suggested to arrange an array of beam splitters in order to
implement a basic non-deterministic gate \cite{erg}. A feasible
linear optical scheme \cite{zjy} was proposed to generate
polarization-entanglement by making use of single-photon quantum
non-demolition measurements based on an atom-cavity system
\cite{pjj}. There is a potential interest in generating
entanglement of optical modes with greater photon numbers.
Entangled $N$-photon states of the form
\begin{eqnarray}
\Psi=\frac{1}{\sqrt{2}}(|0,N\rangle+|N,0\rangle)\label{1}
\end{eqnarray}
are of much interest in respect of the phase sensitivity in a
two-mode interferometer \cite{int}. They should allow a
measurement at the Heisenberg uncertainty limit \cite{djw}.
Recently it was shown that such states allow sub-diffraction
limited lithography \cite{pk}. In the case of $N=2$ the entangled
$N$-photon state (\ref{1}) can be generated easily by using linear
optical elements. For higher values of $N$, a scheme was proposed
by using nonlinear media \cite{ge}. It was assumed that the
generation of quantum states of this type is possible by using
linear optical schemes. Recently, the first linear optical scheme
was proposed to entangle four-photon states \cite{cerf1}.
Furthermore, it was shown that any two-mode photon states can in
principle be generated with a linear optical scheme based on
non-detection \cite{zou,XuBo}. Kok et al presented a linear
optical scheme to generate the state (\ref{1}), which is based on
a $N$-fold
photon coincidence detection.\\
At first we show that two-photon polarization entanglement can be
generated with our scheme, if the initial state is
$|H\rangle_1|H\rangle_2$. We use the abbreviations $H$ and $V$ to
denote the horizontal and vertical linear polarization of each
photon. In Figure \ref{setup} the required symmetric experimental
setup of our protocol is shown. Two optical input modes are
entangled by the beam splitter $BS$. In each of these entangled
branches a polarization rotator, a single-photon injection block
and another rotator is inserted. Finally these branches pass a
polarization beam splitter $PBS$. This setup looks like an optical
interferometer with the possibility to vary the angle of the
rotators continuously, to inject a photon and to get classical
information with the aid of
the detectors $D_1$ and $D_2$.\\

In the following the performance of this arrangement of linear
optical devices will be analyzed in detail. The action of the
symmetric beam splitter $BS$ can be described by the unitary
operator
\begin{eqnarray}
U_{bs}=\exp[\frac{\pi}{4}(a_{1H}a_{2H}^{\dagger}-a_{1H}^{\dagger}a_{2H})]
\,.\label{2a}
\end{eqnarray}
After passing the beam splitter $BS$ the initial state is
transferred into the other mode
\begin{eqnarray}
\Psi_1&=&\frac{1}{\sqrt{2}}(|2H\rangle_1|0\rangle_2-|0\rangle_1|2H\rangle_2)
\,.\label{2}
\end{eqnarray}
The two output modes of the beam splitter pass through two
polarization rotators with the rotation angle $\theta$. These
linear optical devices can be described by the transformation
\begin{eqnarray}
a_H^{\dagger}&\rightarrow &(\cos{\theta})a_H^{\dagger}+(\sin{\theta})a_V^{\dagger};\nonumber\\
a_V^{\dagger}&\rightarrow
&(\cos{\theta})a_V^{\dagger}-(\sin{\theta})a_H^{\dagger}\,.\label{transf}
\end{eqnarray}
In order to obtain the maximum efficiency of our
scheme the rotation angles $\theta$ are calibrated to fulfill
$\cos\theta=\sqrt{\frac{1}{3}}$. The two-photon state transforms
to
\begin{eqnarray}
\Psi_2&=&\frac{\sqrt{2}}{3}[0.5|2H\rangle_{1^{\prime}}|0\rangle_{2^{\prime}}
+|HV\rangle_{1^{\prime}}|0\rangle_{2^{\prime}}+|2V\rangle_{1^{\prime}}|0\rangle_{2^{\prime}}
\nonumber\\
&&-0.5|0\rangle_{1^{\prime}}|2H\rangle_{2^{\prime}}-|0\rangle_{1^{\prime}}
|HV\rangle_{2^{\prime}}-|0\rangle_{1^{\prime}}|2V\rangle_{2^{\prime}}
] \,.\label{3}
\end{eqnarray}
The two output modes of the rotators pass two polarization beam
splitters: $PBS_1$ and $PBS_2$. Since the polarization beam
splitter transmit only the horizontal polarization component and
reflect the vertical component, the state evolves into
\begin{eqnarray}
\Psi_3&=&\frac{\sqrt{2}}{3}[0.5|2H\rangle_{3}+|H\rangle_{3}|V\rangle_{4}+|2V\rangle_{4}
\nonumber\\
&&-0.5|2H\rangle_{5}-|H\rangle_{5}|V\rangle_{6}-|2V\rangle_{6}
]\,. \label{4}
\end{eqnarray}
The scheme uses the output modes $4$ and $6$ to couple separately
a single-photon source as the second input port of the symmetric
beam splitters $BS_1$ and $BS_2$. We assume that these
single-photon sources are in the single-photon state $|1\rangle$.
If the two-fold coincidence detection results in one photon in
each detector $D_1$ and $D_2$ the quantum state is projected into
\begin{eqnarray}
\Psi_4&=&\frac{1}{2}[|2H\rangle_{3}-|2V\rangle_{4^{\prime}}
\nonumber\\
&&-|2H\rangle_{5}+|2V\rangle_{6^{\prime}} ] \,.\label{5}
\end{eqnarray}
Conditioned on this outcome the polarization beam splitters
$PBS_3$ and $PBS_4$ transform the two-photon quantum state into
\begin{eqnarray}
\Psi_5&=&\frac{1}{2}[|2H\rangle_{7}-|2V\rangle_{7}
\nonumber\\
&&-|2H\rangle_{8}+|2V\rangle_{8} ]\,. \label{6}
\end{eqnarray}
These two output modes $7$ and $8$ pass through two polarization
rotators which are calibrated to fulfill:
$\theta_2=\frac{\pi}{4}$. Finally the quantum state
\begin{eqnarray}
\Psi_6&=&\frac{1}{\sqrt{2}}[|HV\rangle_{7^{\prime}}-|HV\rangle_{8^{\prime}}
]\label{7}
\end{eqnarray}
is incident on a polarization beam splitter in order to obtain the
two-photon polarization entangled quantum state
\begin{eqnarray}
\Psi_7=\frac{1}{\sqrt{2}}(|H\rangle|V\rangle-|V\rangle|H\rangle)\,.\label{tme}
\end{eqnarray}
In order to present the principle idea of our scheme to generate
entangled $N$-photon states we introduce in Figure \ref{bb} simple
building blocks $\Theta$ as optical two-mode devices. Two separate
single-photon injections and two photon-number detectors are
involved in this device. In the following we will demonstrate the
generalization capability of the concept to entangle $N$-photon
states. A detailed estimation of the generalized scheme will be
presented in the following. It will be shown how the quantum state
(\ref{1}) can be generated and how different
terms of the $N$-photon state can be deleted.\\
We consider an arbitrary $N$-photon state
\begin{eqnarray}
\Psi_{in}=\sum_{n=0}^{N}C_n|n\rangle_a|N-n\rangle_b\label{psin}
\end{eqnarray}
to be the input mode of the basic block $\Theta$. In contrast to
the previously analyzed two-photon system we label in Figure
\ref{bb} the four input modes $1,2,3,4$ with the letters
$a,b,c,d$. In order to formulate the functionality of this basic
block for arbitrary photon number states transformation we relate
to these modes the annihilation operators $a,b,c,d$ and the
creation operators
$a^{\dagger},b^{\dagger},c^{\dagger},d^{\dagger}$. The interaction
with the beam splitters $BS_1$ and $BS_2$ can be formulated with
the unitary operators:
\begin{eqnarray}
U_1&=&\exp[\theta(ac^{\dagger}-a^{\dagger}c)] \label{8}\\
U_2&=&\exp[\theta(bd^{\dagger}-b^{\dagger}d)]\,. \label{9}
\end{eqnarray}
The output of these beam splitters is the four-mode state
\begin{eqnarray}
\Phi&=&U_1U_2|\Psi_{in}\rangle|1\rangle_c|1\rangle_d =
\sum_{n=0}^{N}\frac{C_n}{\sqrt{n!(N-n)}}(\cos\theta
a^{\dagger}+\sin\theta c^{\dagger})^n(\cos\theta
c^{\dagger}-\sin\theta a^{\dagger})\nonumber\\
&& \times (\cos\theta b^{\dagger}+\sin\theta
d^{\dagger})^{N-n}(\cos\theta d^{\dagger}-\sin\theta
c^{\dagger})|0\rangle_a|0\rangle_b|0\rangle_c|0\rangle_d\,.
\label{10}
\end{eqnarray}
The function of the building block is based on post-selection. A
two-fold coincidence detection projects the four-mode state $\Phi$
into the two-mode states $\Psi$. Conditioned on the coincidence of
one photon in each detector the (not normalized) quantum state
\begin{eqnarray}
\Psi=\sum_{n=0}^{N}C_n\cos^{N-2}(\cos^2\theta -n\sin^2\theta)
(\cos^2\theta -(N-n)\sin^2\theta)|n\rangle_a|N-n\rangle_b
\label{11}
\end{eqnarray}
is generated. We intend to demonstrate that the term
$|i\rangle_a|N-i\rangle_b$ and the term $|N-i\rangle_a|i\rangle_b$
of the input state (\ref{psin}) (with $n=i$ and $n=N-i$) can be
deleted by changing the value $\theta$ of the beam splitters
appropriately. This can be achieved by choosing the parameter
$\theta$ to satisfy
$\tan\theta=\sqrt{i}$.\\
Now we propose the generation of the maximally entangled
$N$-photon state (\ref{1}) by arranging the building blocks
$\Theta$ like it is shown in Figure \ref{nt}. For simplicity, we
consider the $2N$-photon state
$\frac{1}{\sqrt{2}}(|0,2N\rangle+|2N,0\rangle)$, but the same
scheme can also be used to generate the entangled $2N+1$-photon
state $\frac{1}{\sqrt{2}}(|0,2N+1\rangle+|2N+1,0\rangle)$. As the
input mode of the symmetric beam splitter $BS$ we use $N$ photons
in the mode $1$ and $N$ photons in the mode $2$. The output of
this first beam splitter is the entangled $2N$-photon state:
\begin{eqnarray}
\Psi=\frac{1}{{2^N}}\sum_{m=0}^N(-1)^m\frac{\sqrt{(2m)!(2N-2m)!}}
{m!(N-m)!}|2N-2m\rangle_{1^\prime}|2m\rangle_{2^\prime}
\,.\label{8a}
\end{eqnarray}
In order to generate the maximally entangled state (\ref{1}) $M$
basic elements are required $\{M=N/2$, if $N$ is even;
$M=(N-1)/2$, if $N$ is odd$\}$. Conditioned on the $2M$-fold
coincidence detection the state
$\Psi=\frac{1}{\sqrt{2}}(|0,2N\rangle+|2N,0\rangle)$ can be
generated, if the parameters are chosen appropriately:
$\tan\theta_i=\sqrt{2i}$, ($i=1,\cdots,M$). The probability of
this outcome is
$\frac{(2N)![(2N-2)!!]^2[(N-1)!!]^{2N}}{2^{2N-1}(N!)^4(N!!)^{2N}}$,
if $N$ is odd. Otherwise the probability of this outcome will be
$\frac{(2N)![(2N-2)!!]^2[N!!]^{2N}}{2^{2N-1}((N-1)!)^2((N+1)!)^2((N+1)!!)^{2(N)}}$.\\
In order to demonstrate, how entangled $2N+1$-photon states can be
generated, we require that the input of the symmetric beam
splitter is in a quantum state with $2N+1$ photons in the mode $1$
and $0$ photons in the mode $2$. The spatially separated output
photons of the beam splitter are incident on $N$ basic elements
whose parameter is chosen to be $\tan\theta_i=\sqrt{i}$,
($i=1,\cdots,N$). Based on the $2N$-fold photon coincidence
detection, the maximally entangled $2N+1$-photon state (\ref{1})
will be obtained. The probability of success is
$\frac{[(2N)!]^2}{4^{N}(N+1)^{2N+1}[N!(N+1)!]^2}$.\\

In summary, we have suggested a feasible scheme to prepare
polarization-entangled quantum states and entangled $N$-photon
states by using linear optical devices. In the case of the
polarization-entangled two-photon states the probability of the
outcome will be $1/18$. This is a slightly smaller value than
$1/16$, which the scheme with the non-deterministic gate
\cite{erg} makes possible. But the experimental setup of our
scheme is simpler. Instead of four detectors only a two-fold
coincidence detection is required. Thus, the number of detectors
is reduced to the half. In the case of entangled $2N$-photon
states it is shown that not more than $N$ detectors are required,
which is definitely less than other schemes \cite{zou,XuBo,cerf2}
need. An entangled $6$-photon state can be generated with the
probability of $9.7\%$. This probability of the outcome is three
times bigger than that of the scheme \cite{zou} and much
bigger than that of the schemes \cite{XuBo,cerf2}.\\

A generalization to a multi-mode block can be made easily, because
the building blocks, which are introduced with the Figure
\ref{bb}, don't couple the input modes. The multi-photon
coincidence detection requires only the classical information,
which each detector provides. This is a main difference to the
building blocks, which Dowling et al. suggested in their paper
\cite{cerf2}. In our scheme entanglement is only generated in the
first beam splitter, which is shown in Figure \ref{nt}. Thus,
four-mode entangled photon states in the form of
$\frac{1}{2}(|N\rangle|N-m\rangle|0\rangle|0\rangle+|N-m\rangle|N\rangle|
0\rangle|0\rangle+|0\rangle|0\rangle|N\rangle|N-k\rangle+|
0\rangle|0\rangle|N-k\rangle|N\rangle)$ can be generated within
the generalized concept. These quantum states were employed to
create two dimensional patterns on a suitable
substrate in quantum optical lithography \cite{pk}.\\
One of the difficulties of our scheme in respect to an
experimental demonstration consists in the requirement on the
sensitivity of the detectors. Other schemes \cite{cerf1,cerf2},
which are based on a multi-fold coincidence detection, pose the
same requirements on the capability of the detectors. Recently,
the experimental techniques for single-photon detection made
tremendous progress. A photon detector based on a visible light
photon counter was reported, which can distinguish between a
single-photon incidence and the two-photon incidence with high
quantum efficiency, good time resolution and low bit-error rate
\cite{yyy}. Another difficulty is the availability of
photon-number sources. Currently available triggered single-photon
sources operate by means of fluorescence from a single molecule
\cite{cbb} or a single quantum dot \cite{cs,pm} and they exhibit
very good performance. However, in our scheme, in order to
generate entangled photon state, we need a synchronized arrival of
many photons into input ports of many beam splitters, which will
be experimentally challenging. However, the generated
entangled-photon states can act as a kind of valuable source for
quantum computation and quantum communication.

\newpage

\begin{figure}
\caption{This figure shows the required experimental setup to
generate polarization-entangled two-photon states. The
polarization beam splitters ($PBS$, $PBS_i$(i=1,2,3,4)) transmit
$H$-photons and reflect $V$-Photons. Four polarization rotators
$R_i$ (i=1,2,3,4) are required. $BS$ and $BS_i$(i=1,2) denote
symmetric beam splitters. The scheme requires three photon number
detectors $D_1$ and $D_2$.} \label{setup}
\end{figure}

\newpage

\begin{figure}
\caption{The basic element to entangle $N$-photon states. $\theta$
denotes the parameter of the beam splitter.}\label{bb}
\end{figure}

\newpage

\begin{figure}
\caption{If $M$ basic blocks are arranged $N$-photon states can be
entangled in a very efficient way. Therefore a the choice of the
parameter $\theta_i$ and a $2M$-fold coincidence detection is
needed.}\label{nt}
\end{figure}

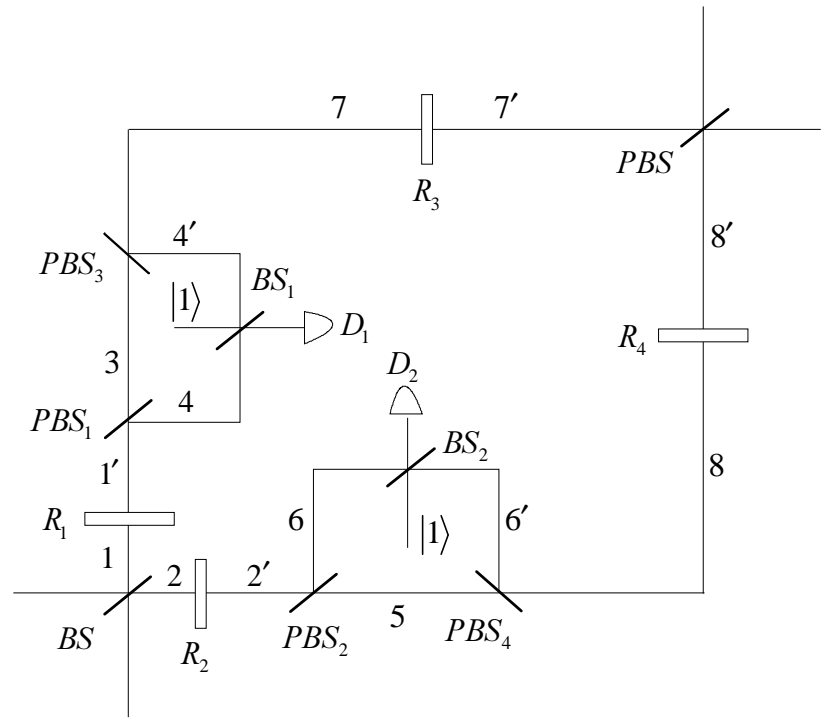
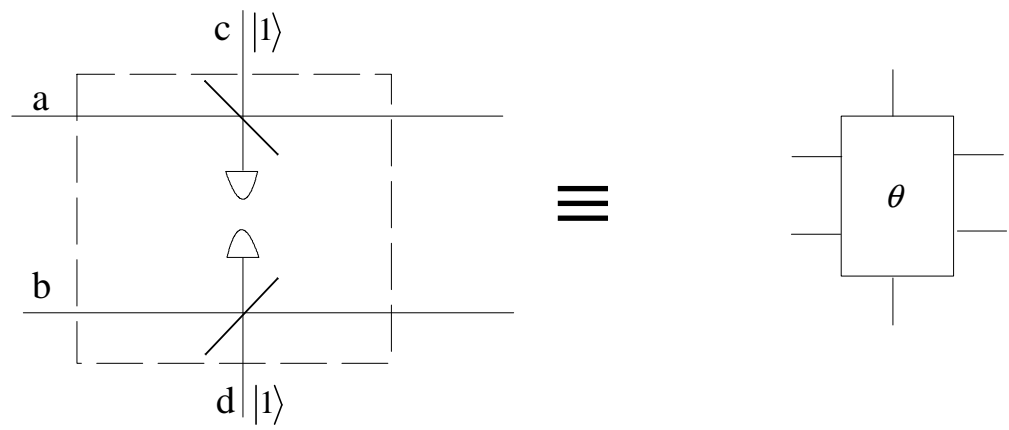
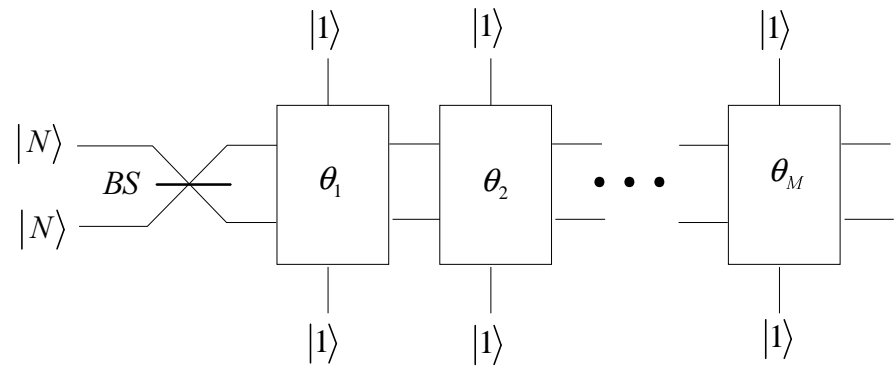
\end{document}